\documentclass[conference]{IEEEtran}
\usepackage{graphicx}
\usepackage{multirow}

\usepackage{amsfonts}
\usepackage[numbers,sort&compress,square]{natbib}
\usepackage{siunitx}
\usepackage{textcomp}
\ifCLASSINFOpdf
\else
\fi
\usepackage{lineno}
\usepackage{nomencl}
\makenomenclature
\usepackage{algorithm}
\usepackage{algorithmic}
\usepackage{amstext}
\usepackage{amsmath}
\usepackage{amsfonts}
\usepackage{color}
\usepackage{epstopdf}
\usepackage{array}
\usepackage[normalem]{ulem}
\usepackage{nomencl}

\usepackage[font=small,skip=0pt]{caption}
\usepackage[numbers,sort&compress,square]{natbib}
\usepackage{subcaption}
\usepackage{graphicx}  
\usepackage{soul}
\usepackage{physics}

\begin{document}

\title{ Optimal Voltage Regulation of Unbalanced Distribution Networks with Coordination of OLTC and PV Generation}

\author{\authorblockN{Changfu Li\authorrefmark{1}, Vahid R. Disfani\authorrefmark{2}, Hamed Valizadeh Haghi\authorrefmark{1} and Jan Kleissl\authorrefmark{1}}
\authorblockA{\\ \authorrefmark{1}Center for Energy Research, University of California San Diego, La Jolla, CA 92093 USA\\
 \authorrefmark{2}Department of Electrical Engineering, University of Tennessee at Chattanooga, TN 37403 USA\\
Emails:  chl447@ucsd.edu, vahid-disfani@utc.edu, valizadeh@ieee.org, jkleissl@ucsd.edu\\ \\}
}


\maketitle
\IEEEpeerreviewmaketitle

\begin{abstract}
Photovoltaic (PV) smart inverters can regulate voltage in distribution systems by modulating reactive power of PV systems. In this paper, an optimization framework for optimal coordination of reactive power injection of smart inverters and tap operations of voltage regulators for multi-phase unbalanced distribution systems is proposed. Optimization objectives are minimization of voltage deviations and tap operations. A novel linearization method convexifies the problem and speeds up the solution. The proposed method is validated against conventional rule-based autonomous voltage regulation (AVR) on the highly-unbalanced IEEE 37 bus test system. Simulation results show that the proposed method estimates feeder voltage accurately, voltage deviation reductions are significant,  over-voltage problems are mitigated, and voltage imbalance is reduced.
\end{abstract}

\section{Introduction}
High shares of photovoltaic (PV) generation present significant challenges to voltage regulation in distribution systems due to power output variability \cite{pecenak2017smart}. Conventional volt-var devices like shunt capacitors are limited in number and slow in response time and therefore unable to regulate feeder voltage during periods of  variable PV generation.

PV smart inverters (SIs) provide an alternative method for regulating voltage \cite{pecenak2017smart}. SI can operate autonomously based on pre-defined Volt/Watt and Volt/VAr curves to regulate voltage \cite{pecenak2017smart}. These autonomous controls are based on local measurements requiring no communications. However, the lack of coordination between SIs and conventional voltage regulation devices can lead to sub-optimal system performance. 

Optimization of SIs considering feeder-wide constraints can achieve optimal power flow  \cite{DallAnese2014,guggilam2016scalable,Robbins2016}. In \cite{DallAnese2014}, semi-definite programming relaxation is leveraged for optimal dispatch of PV active and reactive power. A linear approximation of power flow equations is used in \cite{guggilam2016scalable} for efficiently solving an optimization of PV real and reactive power. An Alternating Direction Method of Multipliers based algorithm is proposed in \cite{Robbins2016} for optimal PV reactive power dispatch and voltage regulation. In these works, cooperation amongst SIs is studied without considering coordination between SIs and other voltage regulation devices like voltage regulators (VRs). Uncoordinated operation of VRs and SIs can cause unintended VR switch operations \cite{kraiczy2017parallel}. PV real and reactive power and VR tap position are optimized simultaneously to minimize voltage deviations in \cite{Christakou2013}. The optimization in \cite{nguyen2018exact} also coordinates PVs, VRs, and shunt capacitors while meeting voltage operation limit constraints.

Non-linear AC power flow constraints render the optimization problem non-convex and computationally intensive for large distribution networks. Different linearization techniques have been applied in the literature to address these concerns. Reference \cite{Christakou2013} derives sensitivity coefficients of node voltages to approximate voltage change as a function of SI real power, reactive power, and VR tap positions. However, the sensitivity of coefficients for VR tap position is calculated assuming that they are located at the substation and the method may not be applicable to distribution feeder with VR in the middle of the feeder. Linearized power flow equations are exploited for improving solution speed in \cite{guggilam2016scalable}. The solution, however, does not coordinate SIs and VRs and is not validated on multi-phase and unbalanced distribution feeders. Further, the voltage estimate from the linear approximation differs substantially from the actual system voltage. SIs, VRs and shunt capacitors are coordinated in \cite{nguyen2018exact} and the method is tested on unbalanced feeders. Since the non-linear terms in power flow constraints are not relaxed, the solution speed is relatively slow. Despite dedicated computational strategies, the average solution speed is 5~s for the small IEEE 34 bus feeder using a high performance PC.

In our previous work \cite{Li2018}, we proposed a linearization technique to convexify the an optimization problem with multiple, coordinated VRs. Voltage violations are mitigated and the method is proven to be computationally efficient. In this paper, we extend \cite{Li2018} by proposing a new linearization method to represent feeder voltage considering coordination between SI reactive power and VR, which is not taken into account in \cite{DallAnese2014,guggilam2016scalable,Robbins2016,Li2018}. By relaxing the non-linear AC power flow constraints, a substantial solution speed-up over \cite{nguyen2018exact} is achieved. Also, VRs do not have to be located at substation as opposed to \cite{Christakou2013}. A sensitivity study shows that the proposed method provides more accurate voltage estimation compared to \cite{guggilam2016scalable}. As each node is modeled independently, the method is compatible with unbalanced feeders as demonstrated on the highly unbalanced IEEE 37 bus test network.

\section{Linearized Model of Voltage Magnitude}
\label{LMTOVI}

\subsection{Linearization of Feeder Nodal Equation}

Consider the feeder nodal voltage equation: 
\begin{align}
V=ZI, 
\label{V-I}
\end{align}
where $V$ is the vector of voltages for all nodes, $I$ is the vector of net node current injections and $Z$ is the feeder impedance matrix. 
A linear approximation of the perturbations in node voltage resulting from changes in impedance and current ($\partial V / \partial (ZI)$) leads to
\begin{align}
\Delta V=\Delta Z\cdot I_0+Z_0 \cdot \Delta I,
\label{VI_derivaive}
\end{align}
where the subscript $(0)$ represents unperturbed parameters.
$\Delta Z$ is a function of tap position changes of VRs and $\Delta I$ is a function of the current injection changes by PVs and loads. Modeling $\Delta V$ requires modeling the effects of VR tap changes on $\Delta Z$ and current source changes on $\Delta I$. 

\subsection{Modeling VR Tap Operation Effects on Voltage}
\label{VoltEffectTO}
VR tap operation effects on voltage can be determined by modeling its effects on $Z$, which is a function of tap ratio $a$. 
 The tap ratio is the ratio of transformer secondary voltage with respect to the primary voltage, which is related to tap position $\tau$ by the linear equation,
\begin{align}
 a=1+\frac{\tau}{\tau_{\max}}(a_{\max}-1),  
 \label{tapratio}
\end{align} 
where $a_{\max}$ is the maximum tap ratio corresponding to maximum tap position $\tau_{\max}$. 

$Y_0$ and $Z_0$ are the admittance and impedance matrices associated with the initial tap ratio $a_0$. $\Delta Y$ is the admittance change due to change of VR tap ratio from $a_0$ to $a$. The corresponding change of the impedance matrix can be expressed as, 
\begin{align}
\Delta Z=-Y_0^{-1}\cdot\Delta Y\cdot Y_0^{-1},
\label{DeltaZ}
\end{align}
which will be used in Eq.~\eqref{VI_derivaive} to determine $\Delta V$.

$\Delta Z$ can be modeled if $\Delta Y$ is known. Considering a VR connected between node $i$ of the primary side and node $j$ of the secondary side, only the elements corresponding to these two nodes in $\Delta Y$ are non-zero:
\begin{align}
\Delta Y_{ii}=(a^2-a_0^2)/z_T, \label{dYii}\\
\Delta Y_{ji}=\Delta Y_{ij}=-(a-a_0)/z_T, \label{dYij} 
\end{align}
where $z_T$ is the equivalent impedance of the transformer on the winding connected to node $i$. Performing Taylor series expansion for $a^2$ around $a_0$, the non-linearity in Eq.~\eqref{dYii} can be removed, yielding a linear expression,
\begin{align}
\Delta Y_{ii} = (2aa_0 - 2a_0^2)/z_T.
\label{dYii_lin}
\end{align}


 More details on the derivation of  Eq.~\eqref{DeltaZ} and the relationship between $Y$ and $a$ can be found in \cite{Li2018}. Although the fixed current assumption is used in \cite{Li2018} to derive Eq.~\eqref{DeltaZ}, the expression is still applicable for this paper without the assumption since both $\Delta Z$ and $\Delta Y$ in Eq.~\eqref{DeltaZ} are direct results of VR tap changes and are only functions of tap positions.

\subsection{Modeling Voltage Impacts of Current Sources}
PVs and loads are current sources. Any change in their injected currents ($\Delta I$) affects the feeder voltage profile as modeled in Eq.~$\eqref{VI_derivaive}$. The power injections of PVs and loads needs to be specified for modeling their current injections.

The power flow equation is,
\begin{align}
S = P+jQ = VI^*,
\label{node_apparent_power1}
\end{align}
where $S$ is the vector of apparent power injection of all nodes on a feeder, $P$ is the vector of the real power and $Q$ is the vector of reactive power injection. $V$ is the voltage vector and $I^*$ is the conjugate of the net current vector. Expressing the parameters as the initial value plus a perturbation, $V$ can be represented as $V = V_0+\Delta V$. Similarly, $I$ can be written as $I = I_0+\Delta I$. Therefore, Eq.~\eqref{node_apparent_power1} can be rewritten as,
\begin{align}
    S = (V_0+\Delta V)(I_0+\Delta I)^*.
    \label{node_apparent_power2}
\end{align}
Eq.~\eqref{node_apparent_power2} sets up the relation between $\Delta I$ and the power injections of PVs and loads.
 
\subsection{Linearization of Real and Reactive Power Injection Constraints}
\label{node_power_constraints}
The power injections of PV and load nodes need to be constrained for representing load consumption and PV production.
Substituting the real and imagery parts of $V_0$, $\Delta V$, $I_0$ and $\Delta I$ into Eq.~\eqref{node_apparent_power2} yields the real and reactive power injection as,   
\begin{align}
P = (V_{d0}+\Delta V_d)(I_{d0}+\Delta I_d)+(V_{q0}+\Delta V_q)(I_{q0}+\Delta I_q), 
\label{realpower2}
\\Q = (V_{q0}+\Delta V_q)(I_{d0}+\Delta I_d)-(V_{d0}+\Delta V_d)(I_{q0}+\Delta I_q). 
\label{reactivepower2}
\end{align}
where $V_0 = V_{d0}+jV_{q0}$, $I_0 = I_{d0}+jI_{q0}$, $\Delta V = \Delta V_d+j\Delta V_q$ and $\Delta I = \Delta I_d+j\Delta I_q$.

The unperturbed variables (subscript $(0)$) are known. The  terms with $\Delta$ symbol are the unknowns to be solved in the optimization. Imposing constraints of $P$ and $Q$ directly would result in products of two unknown optimization parameters (e.g.  $\Delta V_d\Delta I_q$ in $P$) making the problem non-convex. To address this issues, $P$ and $Q$ are linearized and the constraints are implemented using $\Delta P$ and $\Delta Q$.


After linearization, higher order non-convex square terms are dropped to yield
\begin{align}
\Delta P = V_{d0}\Delta I_d+\Delta V_dI_{d0}+V_{q0}\Delta I_q+\Delta V_qI_{q0},\\
\Delta Q = V_{q0}\Delta I_d+\Delta V_qI_{d0}-V_{d0}\Delta I_q-\Delta V_dI_{q0}. 
\end{align}
The higher order non-convex terms constitute the real and reactive power errors $P_{\rm err} = \Delta V_d \Delta I_d+\Delta V_q\Delta I_q$ and $Q_{\rm err} = \Delta V_q\Delta I_d-\Delta V_d\Delta I_d$.

Assuming a constant power load model, the perturbed power injection should remain the same as $P_0$ and $Q_0$.  Therefore the power injections constraints for load nodes are
\begin{align}
    \Delta P=0,
    \label{deltaP_load}
    \\\Delta Q=0.
    \label{deltaQ_load}
\end{align}

Assuming no real power curtailment for PVs, the real power injections of the perturbed PV nodes remains $P_0$. The reactive power injections of PV nodes are limited by the inverter rated power, $|Q|\leq Q_{\rm max}$. $Q_{\rm max} = \sqrt{S^2 - P^2}$ is the maximum available reactive power of the SI, where $S$ is the inverter rated power. After the linearization, the constraints at the PV nodes become
\begin{align}
    \Delta P=0, 
    \label{deltaP_PV}
    \\|\Delta Q| \leq (Q_{max} - Q_0).
    \label{deltaQ_PV1}
\end{align}
We assume the PV operates with unit power factor before perturbations, or $Q_0 = 0$. Therefore the constraint in Eq.~(17) becomes 
\begin{align}
-Q_{\rm max} \leq \Delta Q \leq Q_{\rm max}.
\label{deltaQ_PV2}
\end{align}

\subsection{Linearization of voltage magnitudes}
Node voltages need to be estimated in the optimization to improve the feeder voltage profile. After linearization, the voltage magnitude of an arbitrary node can be calculated using
\begin{align}
|v|=|v_0|+\Delta|v|=|v_0|+|v_0|^{-1}(v_{d_0}\Delta v_d+v_{q_0}\Delta v_q).
\label{vamaglin}
\end{align}
More details regarding the linearization of voltage magnitude can be found in \cite{Li2018}.
This definition for voltage magnitudes of all nodes sets up an affine relation between the voltage magnitude and optimization parameters, which convexifies the optimization problem.

\section{Feeder-Wide VR and PV Optimization}
\label{sec:opt}
\subsection{Optimization Model} 
\subsubsection{Objective Functions}
The first objective function ($J_1$) is the sum of voltage deviations from $1$ p.u. on the feeder during the optimization horizon,
\begin{align}
J_1=\sum_{i=1}^{N} \sum_{t\in T}(||v_i(t)|-1|),
\label{J1}
\end{align}
where $N$ is the total number of nodes on the feeder, $T$ is the set of time steps in the optimization horizon and $|v_i(t)|$ denotes the voltage magnitude of node $i$ at time step $t$. Minimizing $J_1$ achieves a more homogeneous and steady voltage.

The second objective function ($J_2$) counts the number of 
{TO} as,
\begin{align}
J_2=\sum_{p\in P}\sum_{t\in T}|{\tau_{p,t+1}-\tau_{p,t}}|,
\end{align}
where $P$ is the set of all VRs and $\tau_{p,t}$ denotes the tap position of VR $p$ at time step $t$. All tap changes over 
a defined time horizon $T$ are aggregated in $J_2$.

Combining the two objective functions, the final objective of the optimization is
\begin{align}
\min &&J=w_1J_1+w_2J_2,
\label{optimization_problem}
\end{align}
The weighting factors, $w_1$ and $w_2$ balance voltage regulation performance and total TO. Heavy weighting on $J_1$ will improve the voltage profile at the cost of more TO and vice versa. $w_1 = 1$ and $w_2 = 0.05$ will be used in this paper.

\subsubsection{Constraints}
\label{constraint}
\subsubsection*{Power Flow Constraints}
To ensure that the final solution meets the feeder power flow, the linearized power flow (Eq.~\eqref{VI_derivaive}) is an equality constraint. Further, the equality constraint in Eq.~\eqref{vamaglin} relates the voltage magnitudes to the real and imaginary parts of the node voltages. \\
\subsubsection*{VR Constraints are}
\begin{align}
    \tau_{p,t}\in\mathbb{Z},\\ \tau_{p,\min}\le\tau_{p,t}\le\tau_{p,\max},\\
    |\tau_{p,t}-\tau_{p,t-1}|\le \Delta TO_{p,\max},
\end{align}
where $\tau_{p,t}$ denotes the tap position of VR $p$ at time step $t$, $Z$ represents integer numbers. $\tau_{p,\min}$ and $\tau_{p,\max}$ are the minimum and maximum tap positions, respectively. $\Delta TO_{p,\max}$ avoids unrealistic tap operation consider TO delays by limiting the maximum TOs allowed between two consecutive time steps. $\Delta TO_{p,\max}$ is set to 1 TO per 30 sec.

Eq.~\eqref{tapratio} relates VR tap ratio $a$ to VR tap position $\tau$ and is included as an equality constraint. Eq.~\eqref{DeltaZ} is also included as an equality constraint for representing the relationship between $\Delta Z$ and $\Delta Y$. Eq.~\eqref{dYij}-\eqref{dYii_lin} are also included as equality constraints for relating $a$ with admittance change.

\subsubsection*{Node Power Injection Constraints}
As explained in Section~\ref{node_power_constraints}, node power injections need to be constrained for representing PV and load behaviors. For load nodes, Eq.~\eqref{deltaP_load}-\eqref{deltaQ_load} are included as equality constraints. For PV nodes, Eq.~\eqref{deltaP_PV} is included as an equality constraint and Eq.~\eqref{deltaQ_PV2} is included as an inequality constraint.
\subsubsection*{Source Bus Constraints}
For substation nodes, $\Delta V=0$ is imposed as an equality constraint under the assumption of infinity source bus.


\subsection{Implementation and Forecasts}
\begin{figure} 
\centering
\includegraphics[width=0.5\textwidth]{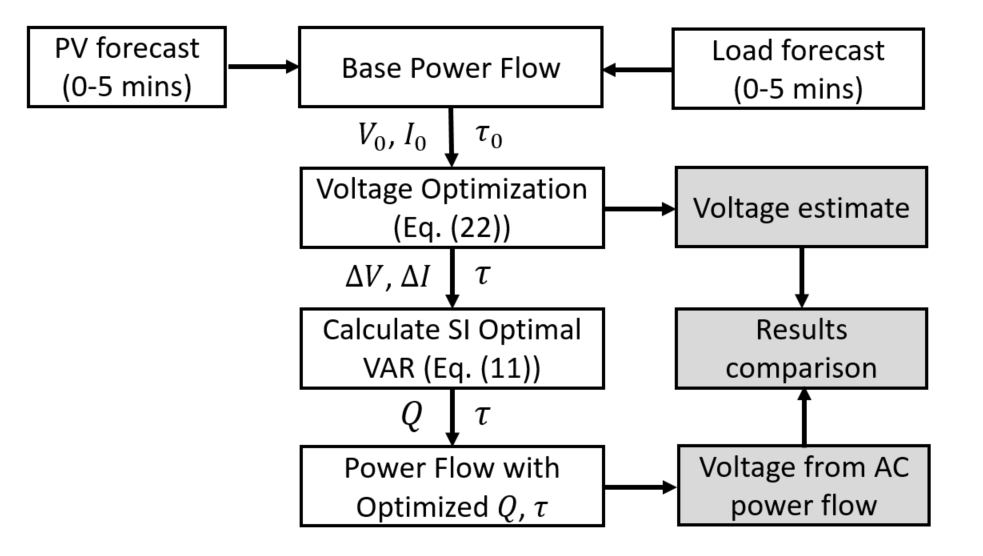}
\caption[Feeder Topologies]{Flowchart of the proposed voltage optimization. PV and load forecasts are used to obtain the linearization voltages and currents ($V_0$, $I_0$) for the next 5 mins from base power flow simulations. Then the voltage optimization per Eq.~\eqref{optimization_problem} is formulated and solved, providing decision values for $\Delta V$, $\Delta I$ and $\tau$. The optimal reactive power of the SI is then calculated using Eq.~\eqref{reactivepower2}. The optimal tap positions $\tau$ and SI reactive power are then input into OpenDSS \cite{OpenDSS} for another power flow. The voltage results from the OpenDSS simulation are then compared to those from the estimation with the proposed method (Eq.~\eqref{vamaglin}).
}
\label{flowchart}
\end{figure}

Fig.~\ref{flowchart} presents the flowchart of the implementation of the proposed voltage optimization. For TOs minimization, the optimization problem is defined over a 5~mins time horizon. $V_0$ and $I_0$ over the next 5 mins are needed for modeling the effects of VR tap position changes and SI reactive power on voltage. In this paper, it is provided from a base power flow run by OpenDSS \cite{OpenDSS} using solar and demand forecasts. 

Sky imagers provide forecasts of PV availability throughout the feeder at high spatio-temporal resolution for the next 5~mins at 30~s resolution \cite{yang2014solar}. A perfect load forecast is assumed using the measured data at the substation provided from the utility. 

\section{Case Study}
\label{casestudy}

\subsection{Distribution Feeder Models}
To evaluate the proposed method, quasi-steady state simulations are carried out on the multi-phase unbalanced IEEE 37 bus feeder. The simulation is performed for 24 hours with 30~sec time step. 30 loads on the feeder result in a ${P_{\rm load\_peak} = 2.73}$~MVA peak demand. 30 PVs with DC power rating ranging from 23.0 to 206.0~kW and totalling $P_{pv\_peak}=4095$~kW  are randomly deployed on the feeder. 10\% oversizing of AC power rating is assumed for the PV interter. The total PV penetration on the feeder is 150\% by capacity:
$PV_{\rm Pen}=\frac{P_{\rm pv\_peak}}{P_{\rm load\_peak}} \times 100\%$.
One VR is installed at the substation. The VR tap position  can vary from -16 to +16 with voltage regulation capability of [0.9 1.1] p.u..

\subsection{Voltage Regulation Methods}
\label{VoltReg}

\subsubsection{Autonomous Voltage Regulation (AVR)} 

The proposed method is benchmarked against the widely-used conventional autonomous voltage regulation scheme (AVR). The two different voltage regulation strategies are summarized in Table.~\ref{VoltRegMethod}. In AVR voltage control devices like VRs operate autonomously based on pre-defined rules without coordination with each other. Only VRs participate in AVR, while PVs do not participate, i.e.  there is no reactive power injection.
VRs change tap to keep the deviation of the local busbar voltage from the preset  reference voltage within certain limits. The VR reference voltage for the test feeder is set to 1.03~p.u. and the voltage regulation bandwidth is 0.0167~p.u.. For better voltage regulation, the tap time delay is set to be 0 sec. All other VR parameters use the default OpenDSS \cite{OpenDSS} values. 

\subsubsection{Optimal Voltage Regulation (OVR)}
\label{OTC}
For OVR, VRs and PVs are coordinated through optimization for voltage deviation reduction as described in Section~\ref{sec:opt}. PV participates in voltage regulation via reactive power absorption and injection. VR tap positions are the outputs of the optimization problem proposed in Section~\ref{sec:opt}. A reference voltage is not needed as the VRs will follow the optimal tap position.

\begin{table}[]
\normalsize
\centering
\caption{Summary of autonomous voltage regulation (AVR) and optimal voltage regulation (OVR).}
\label{VoltRegMethod}
\begin{tabular}{c|c|c}
\hline
                & VR                 & PV            \\ \hline
AVR (benchmark) & Autonomous control & Unity PF      \\ \hline
OVR (proposed)  & Tap optimized      & VAR optimized \\ \hline
\end{tabular}
\end{table}

\section{Distribution Feeder Simulations Results}
\label{result}

\subsection{Voltage Profile}
 




\begin{figure}
\centering
   \begin{subfigure}[b]{0.48\textwidth}
  \includegraphics[width=0.99\textwidth]{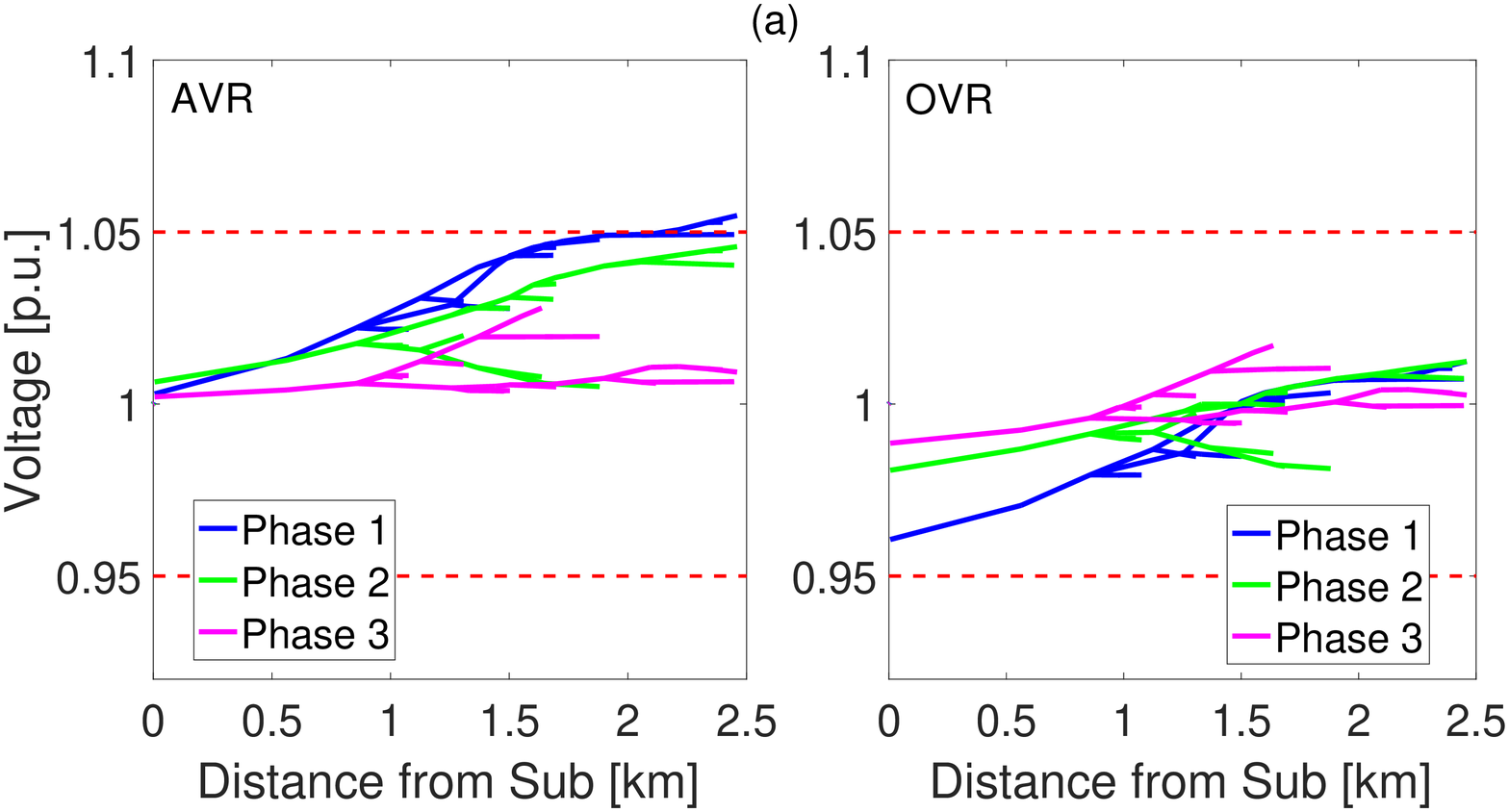}
   \label{Volt12} 
\end{subfigure}

\begin{subfigure}[b]{0.48\textwidth}
\includegraphics[width=0.99\textwidth]{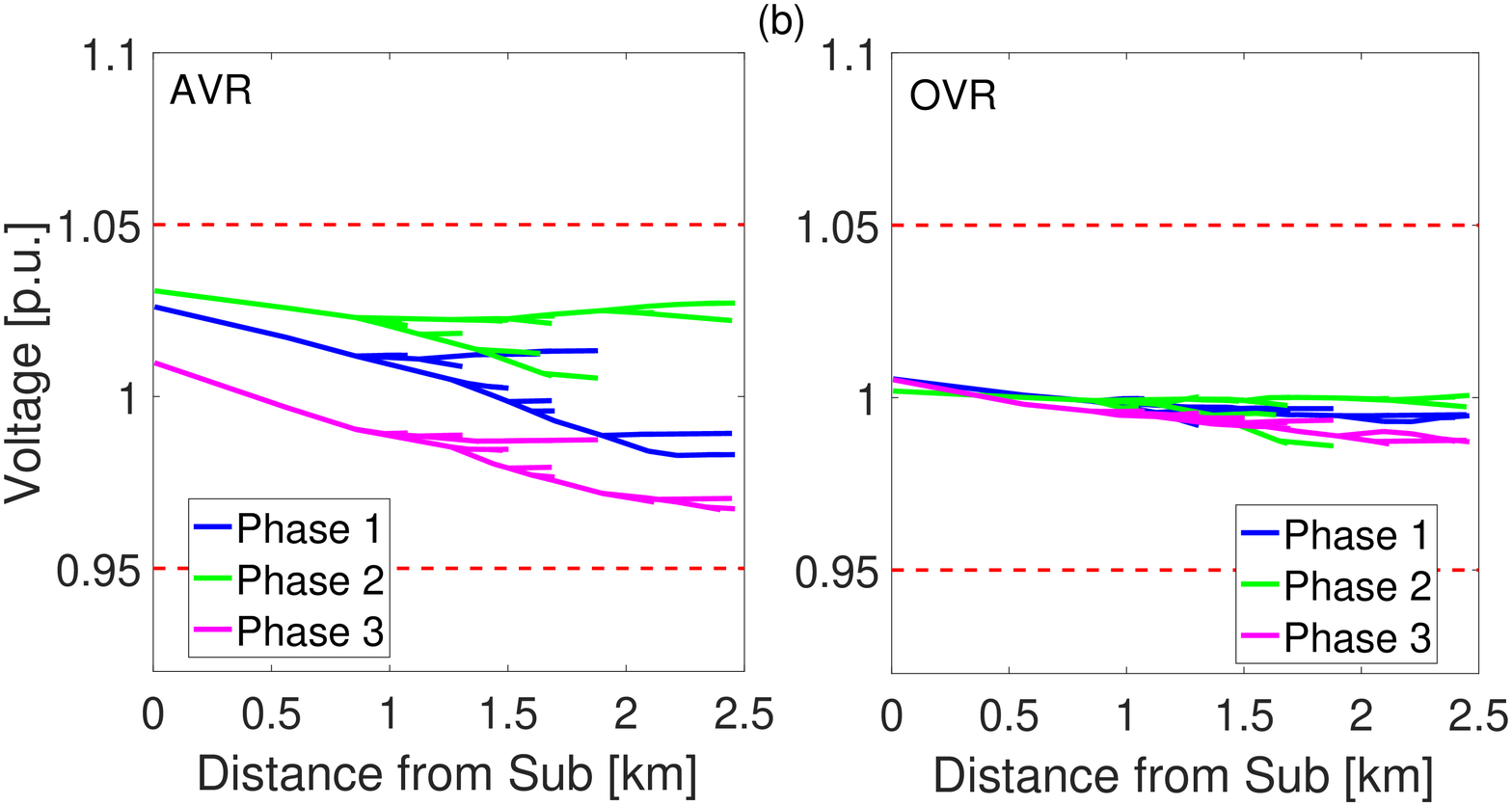}
   \label{Volt21}
\end{subfigure}

\caption[Voltage along the feeder]{Feeder voltage profile at 11:53 (top) and 21:00 (bottom) for the AVR (left) and OVR (right) voltage regulation methods. 
\vspace{-2ex}
\vspace{-0ex}}
\label{voltalongfeeder}
\end{figure}

Fig.~\ref{voltalongfeeder} presents snapshot voltage profiles of the feeder around noon (11:53, medium loading (0.83 MVA), large PV generation) and in the evening (21:00, heavy loading (1.91 MVA), no PV generation). At noon, a voltage increase along the feeder results from reverse power flow caused by excess PV production. For AVR, the voltage increases to 1.055 p.u. at the feeder end, in violation of ANSI standards. The over-voltage violation only occurs on phase 1 and there are large voltage discrepancies across different phases at the feeder end. This indicates significant imbalances on the feeder. On the contrary, OVR manages to keep the voltage of all phases within the [0.95 1.05] p.u. ANSI limits. The voltage imbalance at the feeder end is also reduced since OVR is set up to minimize total voltage deviation, bringing all the voltage closer to 1 p.u.. 

At 21:00 heavy loading causes a large voltage drop with AVR. Again, voltage discrepancies between phases are large: The largest voltage difference occurs between phases 1 and 3 at the feeder end at 0.057~p.u. (equivalent to 57\% of the allowable voltage range). With OVR, the voltages remain close to 1 p.u. across the entire feeder, resulting in a more desirable homogeneous (flat) voltage profile. The voltage imbalance is substantially reduced with a maximum of 0.014 p.u., which is a 75\% reduction compared to the AVR benchmark. OVR squeezes the voltage range on all phases toward 1 p.u. with coordinated reactive power support from PV, reducing the voltage imbalance on the feeder. The strong OVR results are enabled by unlimited reactive power support during night time.

\begin{figure} 
\centering
\includegraphics[width=0.485\textwidth]{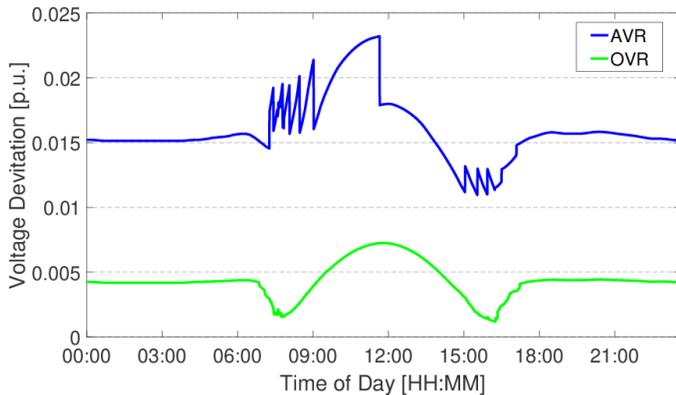}
\caption[]{Time series of mean magnitude of node voltage deviation from $1$~p.u..}
\label{voltdev}
\end{figure}

Fig. \ref{voltdev} compares the average voltage deviation of all nodes on the feeder between AVR and OVR. For AVR, the mean voltage deviation is around 0.015~p.u. during periods without PV production. The voltage deviation increases when PV power production ramps up starting around 08:00 and reaches 0.023 p.u. at noon. With OVR the voltage deviation decreases to below 0.005~p.u. at night and it is always below 0.008 p.u. during the day. The minimum voltage deviations around 08:00 and 16:00 results from PV generation balancing load consumption; therefore minimum power need to be supplied by the substation.

\subsection{Voltage Estimation Accuracy}
\label{acc}
\begin{figure} 
\centering
\includegraphics[width=0.44\textwidth]{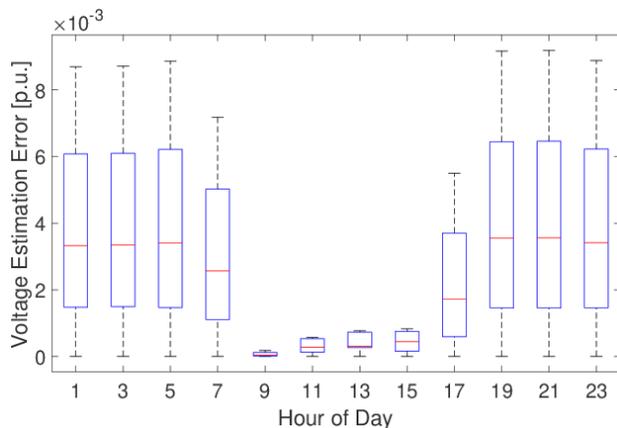}
\caption[]{Distribution of voltage estimation errors. For readability, the results are aggregated every two hours into 12 groups. For example, the box plot of hour 1 is based on the results from 00:00 to 02:00.}
\label{volterr}
\end{figure}

Given that estimated node voltages are used in the formulated optimization (Eq.~\eqref{optimization_problem}) to determine optimal VR tap position and PV reactive power, we examine the errors resulting from the linearization of feeder nodal voltage equations (Eq.~\eqref{VI_derivaive}), admittance matrix (Section~\ref{VoltEffectTO}), power injection constraints (Section~\ref{node_power_constraints}) and voltage magnitude (Eq.~\eqref{vamaglin}). Errors are defined as the differences in estimation of voltage magnitude from Eq.~\eqref{vamaglin} versus the non-linear AC power flow results from OpenDSS: 
\begin{align}E(t)_i = V_{\rm estimate}(t)_i - V_{\rm OpenDSS}(t)_i.
\end{align}
Fig.~\ref{volterr} presents $E(t)$ distributions. Since the estimation match the AC power flow results closely, it can be concluded that the proposed voltage model estimates voltage magnitudes accurately. The maximum error magnitude is 0.009~p.u. and the mean absolute error magnitude is always under 0.004~p.u..


\subsection{Computation Time}

Table~\ref{ct} compares the average computation time in \cite{nguyen2018exact} for the IEEE 34 bus feeder and the IEEE 37 bus feeder using OVR in this paper. Since the non-linear constraints are relaxed in the proposed OVR, the computation time is reduced by 79\%. Solution time is shorter even though \cite{nguyen2018exact} uses a (slightly) smaller feeder and a more powerful computer (Intel Core i7-6700 3.4-GHz processor and 32~GB RAM in \cite{nguyen2018exact} versus  Intel(R) Core(TM) i7-4700MQ 2.8-GHz processor and 16~GB RAM in this paper).

\begin{table}[]
\normalsize
\centering
\caption{Comparison of average computation time (s) per time step in \cite{nguyen2018exact} and OVR case study in this paper.}
\label{ct}
\begin{tabular}{c|c|c|c}
\hline
Test feeder  & \# of nodes & \multicolumn{1}{l|}{Optimization} & Solution time (s) \\ \hline
IEEE 34 & 95          & {[}8{]}                           & 4.55                      \\ \hline
IEEE 37 & 120         & OVR                              & 0.95                      \\ \hline
\end{tabular}
\end{table}

\section{conclusions and Future Work}
\label{Conc}

A novel method of coordinating VRs and PV reactive power for voltage regulation was proposed. OVR is capable of coordination voltage regulation between multiple VRs and PVs. OVR is compared against the conventional AVR through simulations on the highly unbalanced IEEE 37 bus test feeder. 
Results showed the proposed OVR can mitigate over-voltage violations, significantly reduce voltage deviations, and decrease voltage imbalance across phases. This is achieved by effective coordination between VRs and PV reactive power control.

Future work will test the scalability of the proposed method on large real utility distribution feeders. Since OVR relies on solar and power demand forecast to optimize VR tap positions and PV reactive power, future work will also examine the robustness of OVR against forecast errors.  

\bibliographystyle{IEEEtran}
\bibliography{OptTap.bib}

\begin{thebibliography}{10}
\providecommand{\url}[1]{#1}
\csname url@samestyle\endcsname
\providecommand{\newblock}{\relax}
\providecommand{\bibinfo}[2]{#2}
\providecommand{\BIBentrySTDinterwordspacing}{\spaceskip=0pt\relax}
\providecommand{\BIBentryALTinterwordstretchfactor}{4}
\providecommand{\BIBentryALTinterwordspacing}{\spaceskip=\fontdimen2\font plus
\BIBentryALTinterwordstretchfactor\fontdimen3\font minus
  \fontdimen4\font\relax}
\providecommand{\BIBforeignlanguage}[2]{{%
\expandafter\ifx\csname l@#1\endcsname\relax
\typeout{** WARNING: IEEEtran.bst: No hyphenation pattern has been}%
\typeout{** loaded for the language `#1'. Using the pattern for}%
\typeout{** the default language instead.}%
\else
\language=\csname l@#1\endcsname
\fi
#2}}
\providecommand{\BIBdecl}{\relax}
\BIBdecl

\bibitem{pecenak2017smart}
Z.~K. Pecenak, J.~Kleissl, and V.~R. Disfani, ``Smart inverter impacts on
  california distribution feeders with increasing pv penetration: A case
  study,'' in \emph{PESGM, 2017}.\hskip 1em plus 0.5em minus 0.4em\relax IEEE,
  2017, pp. 1--5.

\bibitem{DallAnese2014}
\BIBentryALTinterwordspacing
E.~Dall{\textquotesingle}Anese, S.~V. Dhople, and G.~B. Giannakis, ``Optimal
  dispatch of photovoltaic inverters in residential distribution systems,''
  \emph{{IEEE} Transactions on Sustainable Energy}, vol.~5, no.~2, pp.
  487--497, apr 2014. [Online]. Available:
  \url{https://doi.org/10.1109/tste.2013.2292828}
\BIBentrySTDinterwordspacing

\bibitem{guggilam2016scalable}
S.~S. Guggilam, E.~Dall'Anese, Y.~C. Chen, S.~V. Dhople, and G.~B. Giannakis,
  ``Scalable optimization methods for distribution networks with high pv
  integration.'' \emph{IEEE Trans. Smart Grid}, vol.~7, no.~4, pp. 2061--2070,
  2016.

\bibitem{Robbins2016}
\BIBentryALTinterwordspacing
B.~A. Robbins and A.~D. Dominguez-Garcia, ``Optimal reactive power dispatch for
  voltage regulation in unbalanced distribution systems,'' \emph{{IEEE}
  Transactions on Power Systems}, vol.~31, no.~4, pp. 2903--2913, jul 2016.
  [Online]. Available: \url{https://doi.org/10.1109/tpwrs.2015.2451519}
\BIBentrySTDinterwordspacing

\bibitem{kraiczy2017parallel}
M.~Kraiczy, T.~Stetz, and M.~Braun, ``Parallel operation of transformers with
  on-load tap changer and photovoltaic systems with reactive power control,''
  \emph{IEEE Transactions on Smart Grid}, 2017.

\bibitem{Christakou2013}
\BIBentryALTinterwordspacing
K.~Christakou, J.-Y. LeBoudec, M.~Paolone, and D.-C. Tomozei, ``Efficient
  computation of sensitivity coefficients of node voltages and line currents in
  unbalanced radial electrical distribution networks,'' \emph{{IEEE}
  Transactions on Smart Grid}, vol.~4, no.~2, pp. 741--750, jun 2013. [Online].
  Available: \url{https://doi.org/10.1109/tsg.2012.2221751}
\BIBentrySTDinterwordspacing

\bibitem{nguyen2018exact}
Q.~H. Nguyen, H.~V. Padullaparti, K.~W. Lao, S.~Santoso, X.~Ke, and N.~A.
  Samaan, ``Exact optimal power dispatch in unbalanced distribution systems
  with high pv penetration,'' \emph{IEEE Transactions on Power Systems}, 2018.

\bibitem{Li2018}
\BIBentryALTinterwordspacing
C.~Li, V.~R. Disfani, Z.~K. Pecenak, S.~Mohajeryami, and J.~Kleissl, ``Optimal
  {OLTC} voltage control scheme to enable high solar penetrations,''
  \emph{Electric Power Systems Research}, vol. 160, pp. 318--326, jul 2018.
  [Online]. Available: \url{https://doi.org/10.1016/j.epsr.2018.02.016}
\BIBentrySTDinterwordspacing

\bibitem{OpenDSS}
R.~C. Dugan, ``Reference guide: The open distribution system simulator
  (opendss),'' \emph{Electric Power Research Institute, Inc}, 2012.

\bibitem{yang2014solar}
H.~Yang, B.~Kurtz, D.~Nguyen, B.~Urquhart, C.~W. Chow, M.~Ghonima, and
  J.~Kleissl, ``Solar irradiance forecasting using a ground-based sky imager
  developed at uc san diego,'' \emph{Solar Energy}, vol. 103, pp. 502--524,
  2014.

\end{thebibliography}

\end{document}